\documentclass[aps,prl,superscriptaddress,unsortedaddress,twocolumn,preprintnumbers,amsmath,amssymb]{revtex4-1}
\usepackage{graphicx}
\usepackage{dcolumn}
\usepackage{bm}
\usepackage{upgreek}
\usepackage{txfonts}
\usepackage{color}
\newsavebox{\tempbox}

\begin{document}

\title{Planar Superconductor-Ferromagnet-Superconductor Josephson junctions as scanning probe sensors.}


\author{T. Golod, O.~M. Kapran, and V.~M. Krasnov}
\email{Vladimir.Krasnov@fysik.su.se}

\affiliation{Department of Physics, Stockholm University, AlbaNova
University Center, SE-106\,91 Stockholm, Sweden}

\date{\today }

\begin{abstract}
We propose a novel type of magnetic scanning probe sensor, based
on a single planar Josephson junction with a magnetic barrier. The
planar geometry together with high magnetic permeability of the
barrier helps to focus flux in the junction and thus enhance the
sensitivity of the sensor. As a result, it may outperform equally
sized SQUID both in terms of the magnetic field sensitivity and
the spatial resolution in one scanning direction. We fabricate and
analyze experimentally sensor prototypes with a superparamagnetic
CuNi and a ferromagnetic Ni barrier. We demonstrate that the
planar geometry allows easy miniaturization to nm-scale,
facilitates an effective utilization of the self-field phenomenon
for amplification of sensitivity and a simple implementation of a
control line for feed-back operation in a broad dynamic range.

\end{abstract}

\maketitle

Superconducting Quantum Interference Devices (SQUIDs) are the most
sensitive magnetic sensors, capable of measuring a small fraction
of the flux quantum $\Phi_0=h/2e\simeq 2.07~10^{-15}$ Wb
\cite{Koelle_1999Rev,Kirtley_1999Rev}. In recent two decades SQUID
sensors have been successfully employed in scanning-probe
microscopes with a variety of different designs and operation
modes
\cite{Kirtley_1999Rev,Bouchiat_2009Rev,Granata_2016Rev,Zeldov_2013,Tzalenchuk_2003,Moler_2017}.
The main challenge in scanning SQUID microscopy is a tradeoff
between the field sensitivity and the spatial resolution. Usually
a dc-SQUID, a loop with two junctions, is used for such
applications. For a loop with the area $W\times L$, the magnetic
field sensitivity is determined by the ratio $\Phi_0/WL$ while the
spatial resolution is determined by the loop size and the distance
between the loop and the sample.
For increasing spatial resolution, nano-SQUIDs can be used
\cite{Bouchiat_2009Rev,Granata_2016Rev,Zeldov_2013,Tzalenchuk_2003},
but at the expense of reduced sensitivity. In the conventional
flat geometry of the SQUID, the miniaturization is limited by the
required minimum critical current $I_c \sim 10~\mu$A for avoiding
thermal fluctuations at the operation temperature of several K
\cite{Anticorrelation_2007}. Since $I_c$ scales with the junction
area, this limits the smallest junction size for a given critical
current density and operation temperature. This can be obviated in
a three-dimensional SQUID-on-a-tip geometry \cite{Zeldov_2013}.
The complexity of the full scanning probe SQUID-sensor design
\cite{Kirtley_1999Rev,Moler_2017}, including the SQUID loop with
two junctions and a feed-back loop, is also a hinder for
miniaturization.

In recent years hybrid Superconductor-Ferromagnet (SF) devices are
being actively studied. The competition between superconductivity
and magnetism in SF heterostructures leads to new phenomena and
functionality which is interesting for variety of novel electronic
and spintronic components, \cite{Soloviev_2017} such as phase
shifters
\cite{Feofanov_2010,Golod_PRL2010,Pugach_2011,Birge_2016},
superconducting spin valves
\cite{Khaire_2010,Leksin_2011,Banerjee_2014,Iovan_2014,Blamire_2017,Aarts_2017},
memory cells
\cite{Bolginov_2012,Dresselhaus_2014,Golod_2015,Birge_2018}.

Here we propose a novel type of magnetic scanning probe sensor,
based on a single planar Josephson junction with a magnetic
barrier. We fabricate and analyze experimentally Nb/CuNi/Nb and
Nb/Ni/Nb sensor prototypes with a paramagnetic CuNi and a
ferromagnetic Ni barrier. We show that field sensitivity of such a
sensor may exceed the sensitivity of equally-sized SQUID. This is
caused by a double flux-focusing effect in a planar SFS junction:
first, due to a large demagnetization factor of S-electrodes in a
perpendicular magnetic field, and second, due to a large magnetic
permeability of the F-barrier. Furthermore, we argue that such a
sensor can provide a better than SQUID spatial resolution in one
scanning direction. The two-dimensional planar geometry with just
one junction and no loop allows a flexible design, straightforward
miniaturization to nm-scale and effective utilization of
self-field phenomenon for amplification of sensitivity. We test
scaling of the device down to 200 nm size and suggest and verify
several modifications of the sensor for enhancing its performance,
including usage of several junctions in series for increasing the
readout signal $I_cR_n$ and a control-line for feed-back operation
at the point of maximum sensitivity with a broad dynamic range. We
argue that the combination of high field sensitivity, high spatial
resolution in one direction and simple and flexible 2D design
makes planar SFS junctions beneficial for usage as scanning probe
sensors.

   \begin{figure*}[t]
     \centering
    \includegraphics[width=0.99\textwidth]{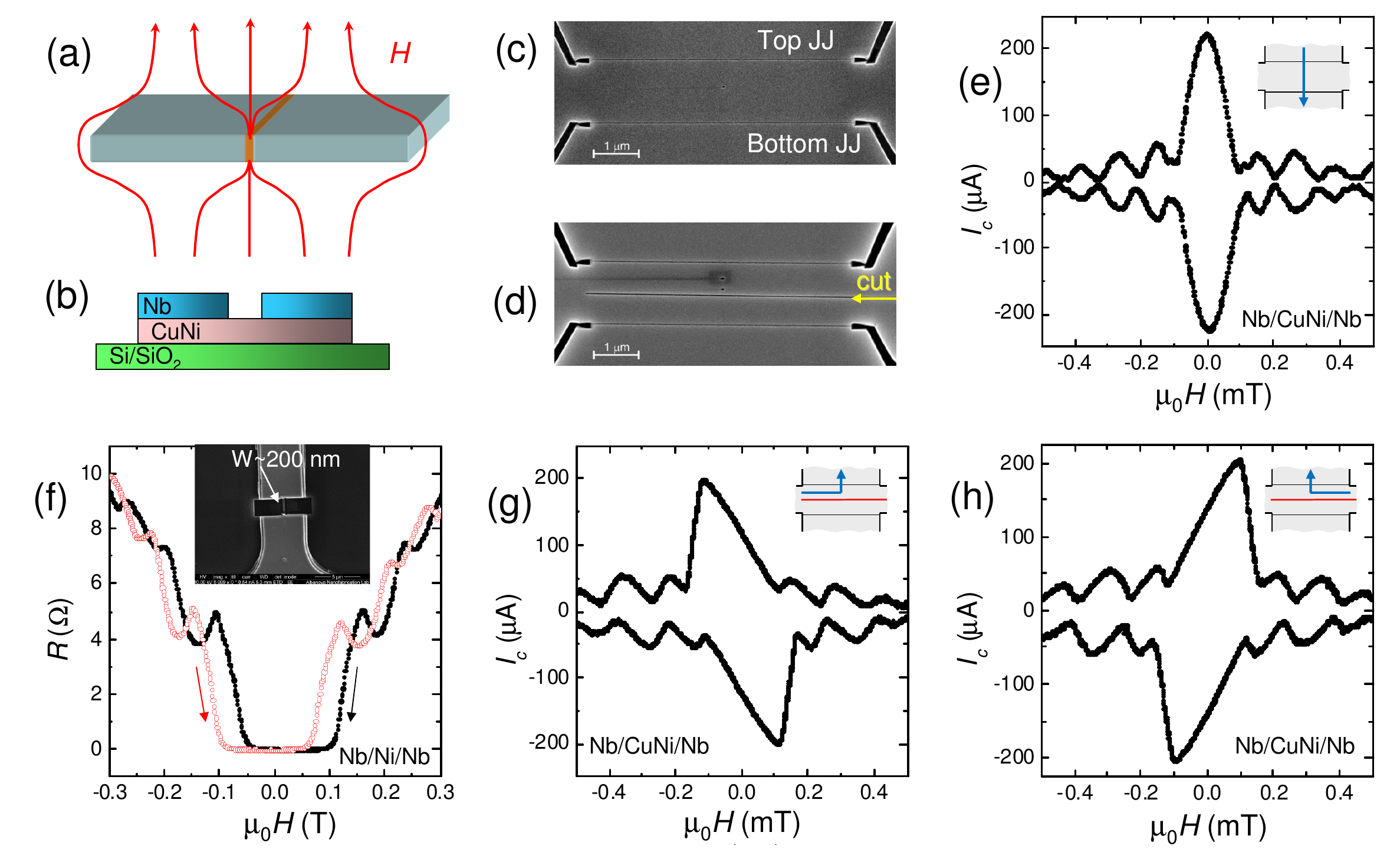}
\caption{
(a) Sketch of a planar SFS junction in perpendicular magnetic
field. Flux focusing in the junction slit occurs due to the large
demagnetization factor of S-electrodes and the large magnetic
permeability of F-barrier. (b) A sketch of crossection of
Nb/CuNi/Nb junction. (c) and (d) SEM images of the studied
Nb/CuNi/Nb sample with two (top and bottom) planar junctions and
four (top, bottom, left, right) electrodes. (c) Before and (d)
after making a separating cut. (e) Fraunhofer-type $I_c$ vs $H$
modulation for the top junction at $T\simeq 6$ K, measured with
uniform bias from top to bottom electrodes, as sketched in the
inset. Absence of hysteresis confirms superparamagnetic state of
CuNi. (f) $R$ vs. $H$ modulation for a $W=200$ nm wide Nb/Ni/Nb
junction. Strong ferromagnetism in Ni leads to the hysteresis in
$R(H)$. Inset shows SEM image of the structure. (g) and (h)
$I_c(H)$ modulation of the top Nb/CuNi/Nb junction with bias
current making a $90^{\circ}$ turn (g) left-to-top and (h)
left-to-top, as indicated in insets. Self-field effect leads to
distortion with a sharp $I_c(H)$ slope at one side. }
    \label{fig1}
\end{figure*}

Planar Josephson junctions (JJs) are formed at the edge between
two superconducting films, see Figure \ref{fig1} (a). They differ
significantly from conventional overlap (sandwich-type) junctions.
In particular, electrodynamics of planar junctions is nonlocal
when electrode thickness is less than the London penetration depth
\cite{Ivanchenko,MintsJLTP1997,Kogan2001,Boris_2013}. When
magnetic field is applied perpendicular to the film plane,
Meissner screening leads to spreading of magnetic field lines
along the surface, as sketched in Fig.~\ref{fig1}(a). This causes
flux-focusing in the junction barrier \cite{Mints2008,Clem2011},
which greatly enhances sensitivity of the planar junction to
magnetic field. The figure of merit for field sensitivity is the
field for achieving a flux quantum. For the symmetric planar
junction with two identical electrodes of sizes $W \times L$
(Width$\times$Length) the quantization field is \cite{Clem2011}
\begin{eqnarray}\label{dH}
  \Delta H\simeq 1.8\frac{\Phi_0}{W^2},~~L>W,\\
  \Delta H\simeq 2\frac{\Phi_0}{WL},~~L<W.
\end{eqnarray}
It has to be compared to the quantization field of the SQUID $\sim
\Phi_0/\langle W \rangle \langle L \rangle$. Note that here
$\langle W\rangle$ and $\langle L\rangle$ are width and length to
the mid-points of the SQUID loop electrodes, while in planar
junctions the full width and length play role. The relative
difference $W,L
>\langle W\rangle,\langle L\rangle$ is increasing with decreasing
sizes. In case of nanoscale sensors this can be an additional
factor in favor of planar junctions. This comparison indicates
that the field sensitivity of planar junctions is comparable
to that for similarly sized SQUIDs. Furthermore, we anticipate
that SFS planar junctions can outperform SQUIDs due to additional
flux focusing in the F-barrier with high magnetic permeability
\cite{Gudoshnikov_2001}. This is the main motivation for our work.

SFS planar junctions are made by focused ion beam (FIB) etching of
a narrow 
grove through an SF bilayer film
\cite{Blamire_1999,PhysicaC_2005}, as sketched in Fig. \ref{fig1}
(b). Details of junction fabrication and characterization can be
found in Refs.
\cite{PhysicaC_2005,Anticorrelation_2007,Golod_PRL2010,Golod_2015}.
The fabrication procedure is simple and reproducible. Junctions
are very stable, no deterioration occurred over five year storage
at ambient conditions.

Fig.~\ref{fig1} (c) represents a scanning electron microscope
(SEM) image of the device, made from Nb/Cu$_{1-x}$Ni$_{x}$
($70/50\,\mathrm{nm}$ bilayer. Thin CuNi films have perpendicular
magnetic anisotropy \cite{Veshchunov_2008}, which in combination
with high magnetic permeability, helps to additionally focus
perpendicular magnetic field in the planar junction and, thus,
further increases the field sensitivity. The superparamagnetic
Cu$_{1-x}$Ni$_{x}$ alloy with Ni concentration close to the
critical for ferromagnetism $x\simeq 0.4$ is chosen to avoid
hysteresis of detector characteristics, which deteriorates the
detector performance \cite{Gudoshnikov_2001}. Furthermore, such
CuNi alloys have a large resistivity, which lifts the junction
resistance to a comfortably measurable value; 

\begin{figure*}[t]
   \centering
    \includegraphics[width=0.99\textwidth]{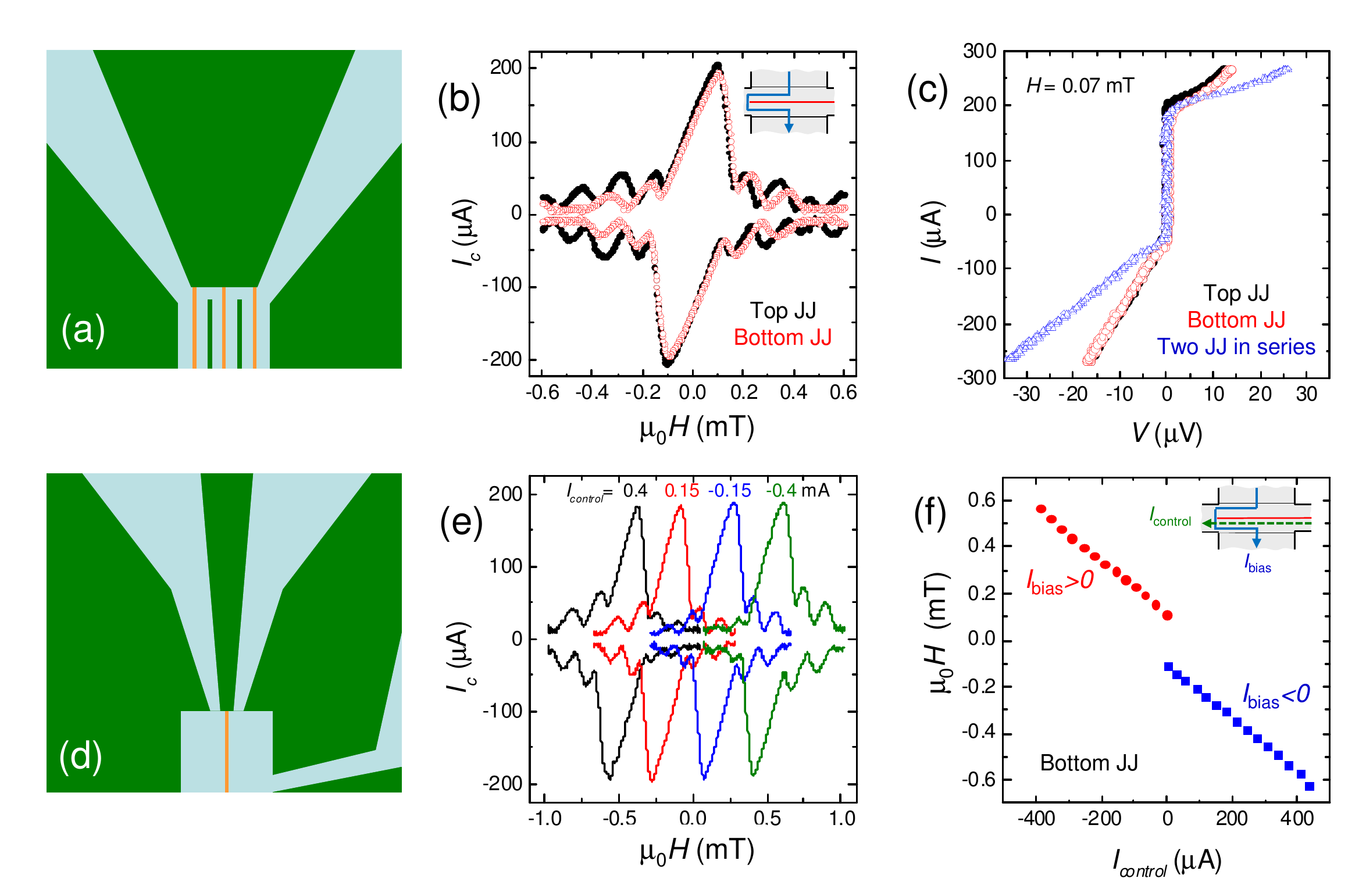}
\caption{
(a) A scanning probe sensor design with three junction in series
(vertical orange lines) and two separating cuts for inducing the
self-field effect. (b) $I_c(H)$ patters for both junctions on
Nb/CuNi/Nb sample with a separating cut. Inset shows the bias
configuration: although the bias current is sent from top to
bottom as in case of Fig.\ref{fig1} (e), the cut leads to current
flow along the junction and thus induces the self-field distortion
of $I_c(H)$. (c) $I$-$V$ characteristics of both Nb/CuNi/Nb
junctions and their serial connection, demonstrating doubling of
the readout voltage. (d) A sensor design with a single junction
and an additional electrode for feedback (control line) operation.
(e) Demonstration of the offset of $I_c(H)$ patterns of the bottom
Nb/CuNi/Nb junction at four control currents along the junction.
Inset in (f) shows the corresponding bias configuration. (f)
Position of the central lobe in $I_c(H)$ for positive and negative
$I_c$ as a function of control current. It represents the dynamic
range for feedback operation $\pm 6$ Oe, corresponding to $\pm 5
\Phi_0$.} \label{fig2}
\end{figure*}

The device in Fig.~\ref{fig1} (c) contains two planar Nb/CuNi/Nb
junctions (marked top and bottom JJ) with similar characteristics.
The junction width is $W \simeq 5.6 \,\mathrm{\muup m}$ and the
separation between them is $L=1.3\,\mathrm{\muup m}$. There are
four long electrodes (top, bottom, left, right) with two bonding
pads for each, allowing independent as well as serial biasing of
the junctions and measurements in the four-probe configuration.
After initial tests, an additional separating FIB cut down to the
substrate was made between the junctions, marked by an arrow in
Fig.~\ref{fig1} (d), to test possible improvements of the sensor.
It is seen that the Nb/CuNi/Nb junctions have dissimilar
electrodes. The outer is much longer than the width and the inner
electrode between the junctions is shorter $L<W$. Each of them
should provide half of the flux from Eqs.(1) and (2),
correspondingly, because those equations were obtained for
symmetric junctions with additive contribution from two equal
electrodes. Therefore the anticipated effective quantization area
$A=\Phi_0/\Delta H$ for our junctions is $\simeq (W^2/1.8 +WL/2)/2
\simeq 10.4~\mu$m$^2$.


Fig. \ref{fig1} (e) shows magnetic field dependence of the
critical current for the top junction (the bottom junction
exhibits similar characteristics). Measurements are done at
$T\simeq 6$ K, before making the separating FIB cut. In this case
the current was sent uniformly through both junctions, straight
from top to bottom electrodes, as sketched in the inset. A regular
Fraunhofer-like modulation indicates good uniformity of JJs. The
$I_c(H)$ does not exhibit hysteresis, confirming that the CuNi
layer is in the superparamagnetic state.
The flux quantization field is $\Delta H \simeq 1$ Oe, yielding
the effective quantization area $A= \Phi_0/\Delta H \simeq
21~\mu$m$^2$.
It is two times larger than the value anticipated from the
junction geometry. We ascribe the corresponding improvement in
sensitivity to additional field-focusing contribution of the CuNi
barrier.

The simple two-dimensional (2D) geometry of a planar sensor with
just one junction allows straightforward miniaturization to
nano-scale. To check the scalability and the effect of
ferromagnetism in the junction, we fabricated a junction with the
width $W\simeq 200$ nm, containing pure Ni, which is a strong
ferromagnet. Inset in Fig. \ref{fig1} (f) shows a SEM image of the
Nb/Ni/Nb junction. The main panel in Fig. \ref{fig1} (f) shows
zero-bias ac-resistance of the junction for upward and downward
sweep of magnetic field at $T\simeq 2$ K. A hysteresis due to
coercivity of ferromagnetic Ni is clearly seen. The $R(H)$
modulation reflects the $I_c(H)$ modulation: minima/maxima of
$R(H)$ correspond to maxima/minima of $I_c(H)$ \cite{Iovan_2014}.
The Fraunhofer-type modulation of $R(H)$ is seen. The period
$\Delta H \simeq 0.9$ kOe is in agreement with Eq. (1) for $W=200$
nm. This confirms that miniaturization of planar junction
detectors to nm scale is indeed simple, but it is accompanied by
the inevitable deterioration of the field sensitivity. It is also
clear that the ferromagnetic barrier introduced the hysteresis in
the detector characteristics, which is detrimental for its
operation \cite{Gudoshnikov_2001}. Therefore, below we will focus
on analysis of micron-sized planar detectors with the
superparamagnetic CuNi barrier.

The 2D planar geometry allows a flexible design and in-situ tuning
of junction characteristics. Figures \ref{fig1} (g) and (h) show
$I_c(H)$ patterns of the top junction with the bias current sent
(g) from the left to the top electrode and (h) from the right to
the top electrode, as sketched in insets. It is seen that the
$I_c(H)$ patterns are distorted compared to the straight uniform
bias case in Fig. \ref{fig1} (e). The distortion is caused by the
self-field effect induced by unevenly distributed bias current
\cite{Vasenko_1981,Barone,Krasnov_1997}. The sign of self-field
depends on the direction of the bias current and reverts between
Figs. \ref{fig1} (g) and (h), demonstrating in-situ tunability of
junction characteristics. Importantly, the self-field effect leads
to a significant sharpening of $I_c(H)$ at one of the slopes of
the central lobe. This greatly enhances the field sensitivity. For
the bias configuration in Figs. \ref{fig1} (g,h) the easy to
detect $1~\mu$A variation of $I_c$ at the sharp slope corresponds
to field and flux variation in mOe and $10^{-3}\Phi_0$ ranges,
respectively.

Our junctions have a rather small $I_cR_n \simeq 20~\mu$V, which
determines the voltage readout. It is possible to tune it in a
broad range by changing the FIB cut depth \cite{PhysicaC_2005}.
Alternatively it can be multiplied by serial arrangement of
several junctions, as sketched in Figure \ref{fig2} (a). In order
to facilitate the self-field in this case we suggest to make
separating cuts between the junctions. In this case the bias
current from left to right electrode will have to flow vertically
along the junction lines, which induces self-field in the
junctions. The studied sample allows to test this idea since it
contains two junctions. Fig. \ref{fig2} (b) shows $I_c(H)$
patterns of both junctions at $T\simeq 6$ K, measured after making
the separating cut, marked in Fig. \ref{fig1} (d). The bias
current is sent from the top to the bottom electrode, just like in
the case of Fig. \ref{fig1} (d), see the sketch in Fig. \ref{fig2}
(b). However, presence of the cut induces the desired self-field
distortion of $I_c(H)$ patterns. Simultaneously it does not
deteriorate significantly the sensitivity, the flux quantization
field $\simeq 1$ Oe. Fig. \ref{fig2} (c) shows corresponding
$I$-$V$ characteristics of individual junction and their serial
sum at $H$ corresponding to maximum positive $I_c$. It is seen
that the $I_cR_n$ of two junctions in series indeed doubles. We
emphasize that such a multiplication of the output is a
consequence of high reproducibility of the fabrication technique,
which leads to nearly identical $I_c(H)$ patterns for the two
junctions, see Fig. \ref{fig2} (b).

The planar geometry allows simple adjustment of the self-field
effect and, thus, the field sensitivity. It increases with
decreasing the angle and the gap between bias electrodes. With
proper design $dI_c/dH$ can be made infinitely large. In Figure
\ref{fig2} (d) we sketch a corresponding simplest design on the
sensor with two almost parallel electrodes and a small gap between
them for enhancement of the self-field. The best sensitivity is
achieved in a narrow field range with the maximum $I_c(H)$ slope.
Therefore, for optimal operation, the sensor should be kept at
this point. In SQUIDS this is achieved by a separate feed-back
loop \cite{Koelle_1999Rev,Moler_2017}. In our planar detector this
can be done in a simpler way by adding one additional electrode,
as sketched in Fig. \ref{fig2} (d). The third electrode attached
to the bottom right side of the junction facilitate current flow
along the junction (from top-right to bottom-right electrode). In
this case the electrode itself acts as a control line for
offsetting the sensor.

The studied Nb/CuNi/Nb sensor has four electrodes, see Fig.
\ref{fig1} (d). Therefore, we can use two of them (top and bottom)
for sending the bias current and the other two (left and right)
for sending the control current, $I_{\text{control}}$. Fig.
\ref{fig2} (e) shows $I_c(H)$ patterns of the bottom Nb/CuNi/Nb
junction measured at four control currents. The bias and control
current configurations are sketched in the inset in Fig.
\ref{fig2} (f). It is seen that the $I_c(H)$ patterns are shifted
by the control current. This allows the feedback operation of the
sensor at the maximum sensitivity point within the field range of
the corresponding shift by $I_{\text{control}}$. Fig. \ref{fig2}
(f) shows the corresponding shift of the maxima at central lobes
$I_c(\Phi=0)$ for positive (circles) and negative (squares) bias
currents as a function of the control current. The linear shift
demonstrates the dynamic range of the feedback operation by the
control current of $\pm 6$ Oe, corresponding to $\pm 5 \Phi_0$ in
the junction for this particular device. This means, that upon
scanning of a sample, the sensor will be able to measure signals
in the broad range $\pm 5 \Phi_0$ in a feedback mode with the
maximum field sensitivity.

Finally, we discuss the spatial resolution of the proposed
scanning probe sensor. In the direction along the junction the
spatial resolution in determined by the size $W$ (electrode
width), similar to SQUID sensors. However in the direction
perpendicular to the junction line, the spatial sensitivity may be
amplified by flux focusing in the narrow barrier with high
magnetic permeability. Therefore, we anticipate that the planar
SFS junction can outperform similar-sized SQUID in terms of
spatial resolution in one scanning direction.

To conclude, we have proposed a novel type of magnetic sensor,
based on a single planar SFS Josephson junction. We have verified
it experimentally using a superparamagnetic CuNi alloy and a
strong ferromagnet Ni as a barrier material. We have shown that a
simple planar geometry facilitates a flexible design of the sensor
and in-situ tuning of its characteristics. This allows utilization
of the self-field effect for significant amplification of
sensitivity, a feed-back operation in a broad dynamic range using
junction electrodes as a control line, and employment of several
junction in series for enhancing the voltage readout. We also
demonstrated a possibility of down-scaling of the device to 200
nm, which, however, inevitably leads to reduction of field
sensitivity, as for small SQUIDs. We concluded that proposed
planar junction scanning probe sensors may outperform
similar-sized SQUIDs both in terms of spatial resolution in the
scanning direction perpendicular to the junction line and in terms
of field sensitivity, due to two flux-focusing effects: a large
demagnetization factor of S-films in perpendicular field and a
large magnetic permeability of the magnetic barrier. We argue that
the combination of high field sensitivity, high spatial resolution
in one direction and simple and flexible 2D design makes planar
SFS junctions beneficial for usage as scanning probe sensors.

\end{document}